\documentclass[aps,prb,twocolumn,groupedaddress]{revtex4}
\usepackage[dvips]{graphicx}

\begin{document}


\title{Magnetotransport properties of individual InAs nanowires}


\author{Sajal~Dhara}
\author{Hari~S.~Solanki}
\author{Vibhor~Singh}
\author{Arjun~Narayanan}
\author{Prajakta~Chaudhari}
\author{Mahesh~Gokhale}
\author{Arnab~Bhattacharya}
\author{Mandar~M.~Deshmukh}
\email[]{deshmukh@tifr.res.in}
\affiliation{Department of Condensed Matter Physics and Materials Science, Tata Institute of Fundamental Research, Homi Bhabha Road,
Mumbai, India 400005}


\date{\today}

\begin{abstract}
We probe the magnetotransport properties of individual InAs nanowires in a field effect transistor geometry.
In the low magnetic field regime we observe magnetoresistance that is well described by the weak localization (WL) description in diffusive conductors. The weak localization correction is modified to weak anti-localization (WAL) as the
gate voltage is increased. We show that the gate voltage can be used to tune the phase coherence length ($l_\phi$) and spin-orbit length ($l_{so}$)  by a factor of $\sim$ 2.
In the high field and low temperature regime we observe the mobility of devices can be modified significantly as a function of magnetic field.
We argue that the role of skipping orbits and the nature of surface scattering is essential in understanding high
field magnetotransport in nanowires.
\end{abstract}


\maketitle



Electron transport in InAs nanowires has been studied extensively \cite{supercurrent1,bjork1,highmobility1,kondosemiconductorNWQD,spinorbitqd} because of interest in high mobility \cite{fetmobility} devices \cite{highmobility1,inasnanowiremobilityref}, low bandgap of 0.35~eV, ability to form ohmic contacts \cite{polysulfidereference}, spin-orbit interaction \cite{bjork1,debaldquantuwires,spinorbitqd} and coherent supercurrent transport \cite{supercurrent1}. Nanowire based devices have been considered to be one of the means to realize spintronic devices like the spin dependent field-effect transistor \cite{dattadas}, and one of the desirable features of such a device is the tunability of spin-orbit interaction \cite{gatecontrolspinorbit}. In order to better understand the spin-orbit interaction and relaxation mechanisms magnetotransport measurements have been done on a network of InAs nanowires  \cite{bjork1} at low magnetic fields. However, studying electron transport in individual nanowires at both low and high magnetic field can provide additional insight into the electron transport in nanowires particularly since there is strong evidence to suggest that in case of InAs there is a subsurface sheet of electrons that participates in the electron transport together with the electrons in the volume \cite{tsui2deg,InAs2deg,inas2deg1}.

 In our experiments we probe the low and high magnetic field electron transport in individual InAs nanowires.
 Our experiments improve upon the earlier work that measured the phase coherence length ($l_\phi$) and spin orbit interaction
 length ($l_{so}$) in a network of nanowires \cite{bjork1}. Our measurements are different because the turn-on threshold of individual nanowires varies ($\sim 5$V) due to the local electrostatic environment and the nature of unintentional doping. We show that the phase coherence length is strong function of gate voltage; as a result ensemble averaging using a network of wires can be a simplification. Our measurements also indicate the tunability of $l_\phi$ and $l_{so}$. The high field magnetotransport measurements with magnetic field perpendicular to the axis of the wire shows that the mobility of the electrons is significantly modified with magnetic field. We argue that degree of surface scattering, together with skipping orbits along the edges play a significant role in determining the variation of mobility at high fields. Our experiments show that the high magnetic field transport provides a way to decouple the contribution of  scattering at surface and or within the volume \cite{alijaveydiameter} -- this is an important challenge in understanding electron and thermal transport in nanostructures.

 \begin{figure}
\includegraphics[width=60mm]{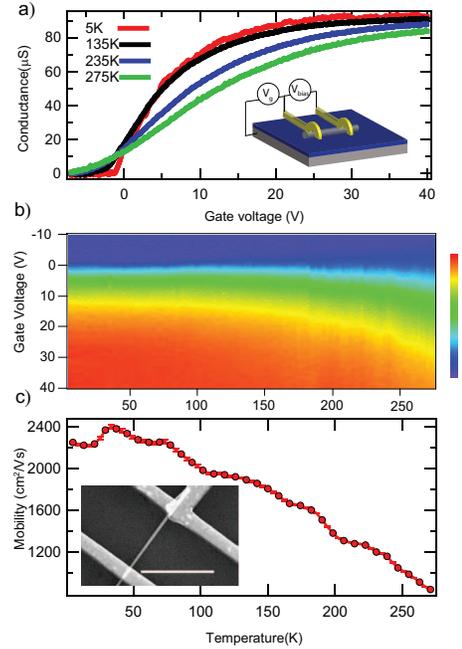}
\caption{\label{figure1}
(color online) (a) Conductance as a function of gate voltage ($V_g$) at various temperatures. The inset shows the device geometry with the nanowires contacted with electrodes on 300~nm thick silicon oxide on Si. The degenerately-doped silicon serves as gate. (b) Colorscale conductance plot of InAs nanowire FET as a function of temperature (4K to 290~K). The minimum (blue) and maximum (red) of the indicated colorscale corresponding to 0 and 97 $\mu$S.
(c) Mobility as a function of temperature. The inset here shows the SEM image with the scale bar of 5$\mu$m. The diameter of the wire is 90~nm.}
\end{figure}

 The InAs nanowires used to make devices in this work were synthesized in a metal organic vapor phase epitaxy system using the vapor-liquid-solid (VLS) technique \cite{NWsynthesislieber,inasVLSgrowth} on a  $< 111 >$ B  oriented GaAs substrate. All the measurements in our work were done on wires 80-90~nm in diameter, 2-5~$\mu m$ long and oriented in the $< 111 >$ direction \cite{bjorkcrystal}. The devices were fabricated by depositing the wires on a degenerately-doped silicon wafer with 300nm of thermally-grown SiO$_2$. The electrodes were then defined by locating the wires relative to a pre-patterned markers using electron beam lithography. After developing, the devices were loaded in a sputtering chamber with an \textit{in-situ} low power plasma etcher. To ensure good ohmic contacts, amorphous oxide / resist residues were removed by in-situ plasma etching before
the deposition of Cr (20~nm) and Au(80~nm) without breaking vacuum. This procedure for  fabricating ohmic contacts differs from the one reported in literature using ammonium polysulphide \cite{polysulfidereference}; however, we have consistently fabricated devices with good ohmic contacts. Fig.~1a shows the conductance (G) of a device as a function of gate voltage ($V_g$) at series of temperatures. The inset shows the schematic of the device and measurement scheme.

 To characterize the devices prior to magnetotransport measurements we performed detailed transport measurements on each device in zero field to study the mobility and on-off characteristics of the device as a function of temperature; these measurements also allow us to confirm the ohmic nature of the contacts. Fig.~1b shows colorscale conductance plot of one such device as a function of temperature from 4~K to 290~K and $V_g$. Fig.~1c shows the plot of mobility as a function of temperature for the device shown in Fig.~1b. The mobility was calculated by taking into account the device geometry and electrical characteristics \cite{mobilityref2,NWcapacitance-saraswat,wucapacitance}.  The variation in the mobility of the device as a function of the temperature shows phonon scattering dominating at higher temperatures and around 30~K the mobility saturates to indicate the residual contribution of impurity scattering.  Using the mobility of the devices at 4~K we can
estimate the meanfree path ($l_e$ $\sim$40~nm) with Fermi wavelength ($\lambda_F \sim$ 30~nm) ( at $V_g=V_{threshold}$~+~5~V). Considering the length of the wire and diameter of wire we find that $l_e < l$, $l$ is the length of wire and $l_e < w$, where $w$ is the diameter of the wire.

\begin{figure}
\includegraphics[width=60mm]{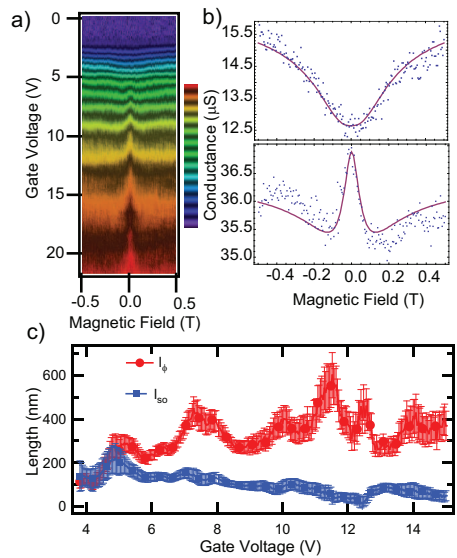}
\caption{\label{figure2}
(color online) a) Colorscale conductance plot as a function of gate and magnetic field at 1.7~K. The banded colorscale is used to accentuate the contours of constant conductance. The weak anti-localization is clearly seen at high gate voltages. The conductance scale spans 0 (blue) to 50 $\mu S$ (red). b) Fits using $\Delta \sigma (B)$ corrections as a function of $B$ to two slices of the data shown in Fig.~2a at $V_g =$ 4.2~V and 8.2~V. c) Phase coherence length $l_\phi$ and $l_{so}$ extracted from fits to the data shown in Fig.~2a.  The diameter and length of the nanowire is 90~nm and 3~$\mu$m respectively.
}
\end{figure}

We next consider the magnetotransport measurements for the nanowires at low magnetic fields as the density of
carriers is modified over a large range by tuning the gate voltage from off to on state. In the data shown we vary the density of electrons upto $\sim$ 5~$\times10^{17}$ cm$^{-3}$.
Low magnetic field transport measurements measuring the variation of conductance as a function of magnetic field allow us to measure the phase coherence length $l_\phi$ and spin-orbit relaxation length $l_{so}$ \cite{altshuler1,weaklocalization1,montambaux-book}. The increase beyond the Drude conductivity when an electron gas is subjected to low magnetic field is known as Weak Localization (WL) and is due to destruction of the constructive interference between time-reversed trajectories; in case of strong spin-orbit interaction this correction becomes negative leading to a drop in conductance known as Weak Anti-Localization (WAL) \cite{montambaux-book}. Fig.~2a shows the color scale conductance plot of one such device as a function of $ V_g$ and $B$. The banded colorscale accentuates the transition  WL to weak WAL regimes
that exhibit an increase in conductance and decrease in conductance respectively as a function of magnetic field
\cite{altshuler1,weaklocalization1,montambaux-book}. The conductance can be fitted based on the calculation of conductance of a nanowire of
length $l$ and diameter $w$ with the magnetic field oriented perpendicular to the direction of current flow.
In our devices $l_e < w$ so we use this \cite{bjork1,altshuler1,weaklocalization1} limit to analyze our data. The fits for the data
are shown in Fig.~2b at two different regimes, namely the WL and WAL, in the same device. The fitting is based on the correction to conductivity, $\Delta \sigma (B) \propto [\frac{3}{2}(\frac{1}{l_\phi^2}+ \frac{w^2 e^2 B^2}{3\hbar^2}+ \frac{4}{3l_{so}^2})^{-1/2} - \frac{1}{2}(\frac{1}{l_\phi^2}+ \frac{w^2 e^2 B^2}{3\hbar^2})^{-1/2}]$,
in the presence of magnetic field $B$ for a rectangular wire of cross-section $w$ and $e$ being the charge of an electron.

\begin{figure}
\includegraphics[width=60mm]{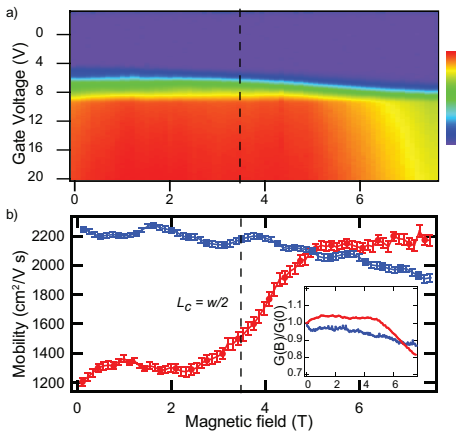}
\caption{\label{figure3}
(color online) a) Colorscale conductance plot of a device at high magnetic field showing the turning on and off of the nanowire FET as a function of magnetic field. The minimum and maximum of the scale correspond to 0 (blue) and 11 $\mu$S (red) respectively. b) Plot of mobility  as a function of magnetic field for two devices
at 1.7~K. The data plotted in red corresponds to device for which the colorscale conductance plot is shown in Fig.~3a. Inset shows the plot of on-state conductance as a function of $B$ for the two devices shown in Fig.3b.}
\end{figure}

We have performed similar measurements on several other devices and qualitative features of the key analysis and features are the same and we discuss them in further detail. Fig.~2c shows the evolution of the phase coherence length ($l_\phi$) and spin-orbit length ($l_{so}$) -- a measure of the spin-orbit interaction within the wire. We observe that there is a continuous variation as a function of gate voltage. We observe peak-like features in the extracted values of $l_\phi$, however, we do not understand the origin of these features. One other interesting feature of our data is that there is a variation in $l_\phi$ and $l_{so}$ even after the on-state conductance saturates. We speculate that this could be a signature of electric field induced change in the spin-orbit interaction. Further experiments with self-aligned gates that subject the FET channel to external electric field are required to better understand some of these observations.

In our analysis we find that both $l_\phi$ and $l_{so}$ are both tuned by $V_g$ by a factor of two. As the turn-on threshold of various nanowire FET devices varies due to the local electrostatic environment the ensemble averaging over a network \cite{bjork1} may suppress mesoscopic effects. The origin of the variation in magnitude of $l_\phi$ and $l_{so}$ could be due to a variety of reasons, including Dresselhaus effect \cite{dresselhausrel}, the Elliot-Yafet mechanism \cite{elliotrel,bjork1} and Rashba effect \cite{rashbaold,rashbanew}. Dresselhaus effect \cite{dresselhausrel}, which depends on bulk induced asymmetry \cite{rashbanew}, and should not contribute because of the absence of spin-splitting in the direction of transport ($< 111 >$). Hansen \emph{et al.} \cite{bjork1} have estimated relaxation due to Elliot-Yafet mechanism \cite{elliotrel,elliotrel2} to be $l_{so} \sim$ 0.5 to 2 $\mu$~m  for similar nanowires. One can estimate for the case of Rashba effect $l_{so}=\hbar^2(m^*eE \alpha_0)^{-1}$ which can arise due to electric field ($E$) across the width of the wire due to the gate, and for bulk InAs \cite{inasbulkconstants} the effective mass $m^*=0.023m_e$ and $\alpha_0= 117 \AA^2$. If the $l_{so}$ varies between 200 to 100 nm as observed in our experiments this would imply a voltage drop of $\sim1.2 $ V as $eE \alpha_0 \sim 10^{-11} eV/m$ across the wire. However, experiments with bulk InAs in MOSFET geometry \cite{rashbainInAsbulk} have observed values $eE \alpha_0 \sim 10^{-11} eV/m$; such values have been observed in heterostructures \cite{gatecontrolspinorbit} as well.

The estimate of $ l_{so}$ from the latter two mechanisms using bulk values is more than the observed $l_{so}$. The work on InAs MOSFETs \cite{rashbainInAsbulk} however clearly shows
the crucial role of electrons at the interface of SiO$_2$ and InAs \cite{tsui2deg,InAs2deg}, particularly at low density of electrons (low V$_g$);this also broadly explain trends seen in our data. There are possibly two mechanisms, one dominating at low V$_g$ corresponding to $l_{so}$ and $\Delta_{so}$ for a 2D electron gas at the surface of nanowire and the other mechanism that takes effect when the electron density in the interior of the wire rises to affect the contribution of electrons at the InAs and SiO$_2$ interface. Further experiments on InAs nanowires that are freely suspended are needed to conclusively resolve the origin of the mechanism.

 We next consider the high magnetic field transport in these nanowires. In the high magnetic field regime when the width of the wire is comparable to the magnetic length ($l_B = \sqrt{\hbar / e B }$) which leads to a modification of the sub-band structure \cite{debaldquantuwires} and could be of interest for studying spintronic systems. Additional effects of confinement become important when the diameter of the radius of cyclotron orbits for the electrons becomes smaller than the radius of the nanowire ($l_c=\hbar k_F/e B < w/2 $).
We have probed high magnetic field transport in InAs nanowires to study the variation in magnetoconductance \cite{magnetoresistancenw,magnetoconductanceosc-tserkovnyak} and mobility of electrons.  In these measurements the magnetic field is oriented perpendicular to the axis of the nanowires.  Fig.~3a shows the plot of conductance of a device as a function of $B$ and $V_g$ at 1.7~K. We point out main features of  the data shown in Fig.~3a -- firstly, the slope $G$ from off-state to on-state varies as a function of $B$, secondly, we see that the on-state conductance $G$ reduces as a function of $B$ and lastly, the change in the $V_{threshold}$ as a function of $B$. Now, we discuss the first observation -- the slope $G$ when the NWFET turns on. Slope of $G$ is proportional to the mobility $\mu$ of the transistor via transconductance $g_m$. Fig.~3b shows plots of $\mu$ as a function of $B$ from two distinct devices.

 The plot of $\mu$ \cite{inasnanowiremobilityref,mobilityref2} as a function of $B$ for the data shown in
 Fig.~3a is depicted in red with open circles. The mobility increases significantly from a value of 1200~cm$^2$V$^{-1}$s$^{-1}$ at 0~T to about 2200~cm$^2$V$^{-1}$s$^{-1}$ at 7.5~T. Fig.~3b also shows the plot of $\mu$ as function of magnetic field for a second device (marked blue with square symbols) for which the mobility exhibits a significant decrease, contrary to what is seen in first device (marked red with circular symbols). In order to understand the microscopic origin of these observations we have also performed measurements on similar devices with magnetic field pointing along and perpendicular to the axis of the nanowire. We find the variation in conductance to be larger when the magnetic field is pointed perpendicular to the axis of the nanowire. We must note that there are device to device variations.

Magnetoconductance variation at high magnetic field could be due to the variation of the contact resistance with magnetic field due to spin-selective scattering at a non-ohmic contact; however, we have not used any magnetic materials for fabricating the contact and the temperature evolution of conductance (Fig.~1a) indicates that the contacts are ohmic. Our observation that the variation in $\mu$ as a function of $B$ is dependent on the relative orientation of the axis of the wire and magnetic field indicates that geometrical effects of electron trajectory are important; this has been observed by others as well \cite{beenaker-review,thornton-roukes-scattering,circularorbitsimaging,skippingorbitmetals}. At zero magnetic field the electron traverses from one electrode to another via a series of scattering events within the volume of the sample and the surface. In our measurements we find that $l > l_e$ and as one increases the magnetic field another lengthscale, $l_c$, becomes relevant; the cyclotron orbit radius -- $l_c=\hbar k_F/e B$. When $l_c < w/2$, the contribution of surface scattering can reduce leading to an increase in the mobility \cite{bismuthnanowiremagnetoresistance}. For the data shown in Fig.~3a the $l_c= \frac{1.4 \times 10^{-7}}{B(T)}$~m and the crossover field (for $w=80$~nm) occurs at a value of $ 3.6 $~T. The concomitant change in the $V_{threshold}$ can also be understood due to the change in the screening of gate voltage once the electrons are confined tightly to the surface and there are localized orbits within the volume of the nanowires \cite{beenaker-review} while $l_c << w/2$. This qualitative picture
changes if the nature of surface scattering is different.
Device to device variations result in different trends in
the dependence of $\mu$ with $B$, as seen in the data from
 the two devices in Fig. 3b.  The main difference between these two devices is their $\mu$ at $0$T. The drop in $\mu$ as a function of $B$ is something that needs further detailed analysis beyond the scope of our manuscript. The unusual role of surface scattering is also seen in another aspect of our data -- the evolution of the conductance at a fixed electron density as a function of magnetic field (shown in inset Fig.3b). We observe that in all our devices the conductance reduces at high magnetic fields and for the case of narrow channels with diffuse scattering negative magnetoresistance is expected \cite{beenaker-review}. If one considers only bulk scattering then this could also happen  when $l_c < l_e $ and in our devices this will occur at $B\sim 3.6$~T; also when $l_c < w/2$ crossover occurs. Surface roughness effects have been theoretically shown to be important \cite{semiconductingNWroughness} in determining transport length scales and may need to be considered. It is also essential to understand the role of the layer of electrons confined close to the surface of InAs nanowires \cite{tsui2deg,InAs2deg,inas2deg1} as this could affect the nature of surface scattering.

  We have described detailed magnetotransport measurements on individual InAs nanowires.
  We find that the $l_{so}$ can be tuned by over a factor of $\sim 2$. One possible explanation of  this
  tunability could be the transition from low to high density of electrons changing the relative contribution of
  electrons in the volume and those confined to the surface of InAs \cite{tsui2deg,InAs2deg}. In the high magnetic
  field regime we find that transition in mobility occurs when the cyclotron orbit size becomes smaller than the width
  of the wire. These measurements also indicate that surface scattering plays a crucial role and magnetic field can be
  used to tune the contribution of two scattering mechanisms in nanostructures.

\begin{acknowledgments}
We acknowledge help of Vikram~Tripathi, Shamashis Sengupta and V.~Priya during the course of this work.  This work was supported by Government of India.
\end{acknowledgments}


\begin{thebibliography}{99}
\bibitem{fetmobility}
We refer to field effect mobility as mobility in our work.
\end{thebibliography}

\end{document}